\documentclass{article}
\usepackage{spconf,amsmath,graphicx}
\usepackage{booktabs}  
\usepackage{multirow}
\usepackage{url}
\usepackage{bm} 
\usepackage{color}
\usepackage{hyperref}

\title{Diagnosing Colorectal Polyps in the Wild with Capsule Networks}

\name{Rodney LaLonde$^{\star}$ \quad 
Pujan Kandel$^{\dagger}$ \quad 
Concetto Spampinato$^{\S}$ \quad 
Michael B. Wallace$^{\dagger}$ \quad 
Ulas Bagci$^{\star}$}

\address{$^{\star}$Center for Research in Computer Vision, University of Central Florida, USA \\
    $^{\dagger}$Division of Gastroenterology and Hepatology, Mayo Clinic, Jacksonville, FL, USA. \\
    $^{\S}$Department of Electrical, Electronics and Computer Engineering, University of Catania, Italy.}

\begin{document}
%
\maketitle
\begin{abstract}
Colorectal cancer, largely arising from precursor lesions called polyps, remains one of the leading causes of cancer-related death worldwide. Current clinical standards require the resection and histopathological analysis of polyps due to test accuracy and sensitivity of optical biopsy methods falling substantially below recommended levels. In this study, we design a novel capsule network architecture (D-Caps) to improve the viability of optical biopsy of colorectal polyps. Our proposed method introduces several technical novelties including a novel capsule architecture with a capsule-average pooling (CAP) method to improve efficiency in large-scale image classification. We demonstrate improved results over the previous state-of-the-art convolutional neural network (CNN) approach by as much as $\bm{43}$\textbf{\%}. This work provides an important benchmark on the new Mayo Polyp dataset, a significantly more challenging and larger dataset than previous polyp studies, with results stratified across all available categories, imaging devices and modalities, and focus modes to promote future direction into AI-driven colorectal cancer screening systems. Code is publicly available at \href{https://github.com/lalonderodney/D-Caps}{https://github.com/lalonderodney/D-Caps}.
\end{abstract}

\section{Introduction} \label{sec:intro}
Among all cancer types, colorectal cancer remains one of the leading causes of cancer-related death worldwide, with the lifetime risk of developing colorectal cancer around $1$ in $23$ in the United States, accounting for roughly $10\%$ of all cases across genders \cite{acs}. The gold standard for colorectal cancer diagnosis is based on the biopsy of colon polyps found during screening (colonoscopy). Due to the vast majority of colorectal cancer cases arising from precursor lesions, referred to as polyps, the identification and resection of pre-malignant polyps during colonoscopy has been shown to decrease colorectal cancer incidence by $40$ -- $60\%$ \cite{Brenner}. However, small and diminutive polyps make up over $90\%$ of polyps detected, with less than half of these classified as pre-malignant, making diagnosis through `optical biopsy' by colonoscopists difficult. 

\begin{figure}[t]
\begin{center}
   \includegraphics[width=0.87\linewidth]{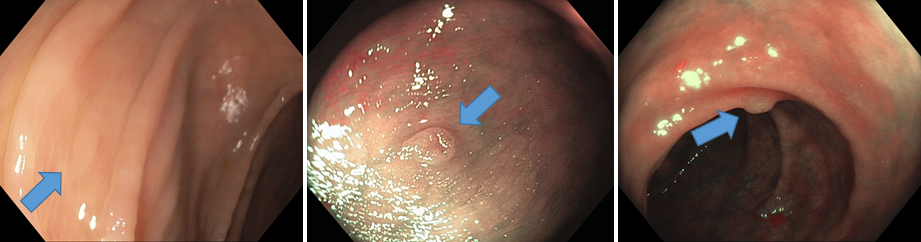}
\end{center}
   \caption{Typical cases on real-world (`in-the-wild') polyp diagnosis cases from the Mayo Polyp dataset. Left to right: hyperplastic, serrated, and adenoma, marked by blue arrows.}
\label{fig:Polyps}
\end{figure}

Colorectal polyps are typically classified into one of three categories: hyperplastic, serrated (comprised of sessile serrated adenomas and traditional serrated adenomas), and adenomas. Example polyps can be seen in Fig.~\ref{fig:Polyps}. Serrated polyps and adenomas are considered premalignant and should be resected during colonoscopy, while hyperplastic polyps are considered benign and can safely be left \emph{in situ}. Unfortunately, existing optical biopsy techniques, cannot currently be recommended in routine clinical practice due to test accuracy and sensitivity falling substantially below recommended levels \cite{Rees}. Therefore, current standards require taking a sample of the polyp and performing histopathological analysis, a somewhat time-consuming and expensive process. Further, performing polypectomies (i.e., biopsy) on non-premalignant polyps is unnecessary, increases procedure-related risks such as perforation and bleeding, and increases procedure-related costs including the cost of histological analysis for diagnosis. Improvements in colonoscopy and optical biopsy techniques have been developed \cite{IJspeert963, Sano}; however, with increased colonoscopy use causing an increase in detected polyps, expecting endoscopists to perform optical diagnosis during colonoscopy screenings might prove too time-consuming to manage in routine clinical practice. There is a high-expectation for artificial intelligence (AI), particularly deep learning, approaches to be adopted into clinical settings for earlier and more accurate diagnosis of cancers.

\textbf{Research Gap:} Previous academic works have achieved remarkable success in this difficult task, with accuracy scores just exceeding $90\%$ \cite{CHEN2018568, Zhang}. \textbf{However, these methods have been applied to academic datasets which are highly unrealistic compared to a `real-world' clinical setting.} For example, the most popular dataset in the literature is the ISIT-UMR Multimodal classification dataset \cite{Mesejo}, containing only $76$ polyps. Each polyp is recorded up-close for approximately 30 seconds (nearly 800 videos frames) from multiple angles, modalities, and focus modes. Such time-consuming and ideal videos cannot be expected in more realistic `in the wild' (i.e., real-world) clinical settings. To address this discrepancy between ideal academic datasets and real-world examples, we performed experiments on the significantly more challenging Mayo Polyp classification dataset, collected at the Mayo Clinic, Jacksonville by \cite{WALLACE20141072} with institutional review board approval. A total of $963$ polyps from $552$ patients were collected, where one image per imaging type of each polyp are chosen by expert interpreters. This dataset is extremely challenging, having only \textbf{single images per imaging mode per polyp}, large inter-polyp variation (e.g., scale, skew, illumination), and often only a single imaging mode provided, while also containing far more polyps collected from more patients than all previous AI-driven diagnosis studies in this area.

To accomplish our task and improve the viability of optical biopsy of colorectal polyps, we design a novel capsule network (D-Caps). Capsule networks provide equivariance to affine transformations on the input through encoding orientation information in vectorized feature representations, and \textit{we hypothesize that a capsule network can better model the high intra-class variation present in the Mayo Polyp dataset and provide superior results to a deep CNN}. Our method introduces several technical novelties including (i) a novel deep capsule network architecture based on the locally-constrained routing introduced in \cite{lalonde2018capsules}, (ii) a capsule-average pooling (CAP) technique which allows us to perform classification on large image sizes, where the original fully-connected capsules of \cite{Sabour} are far too computationally expensive to fit in GPU memory, and (iii) improves the results over CNNs such as Inceptionv3 (Iv3) \cite{szegedy2016rethinking} employed the previous state-of-the-art \cite{CHEN2018568} by a significant margin, while also reducing the amount of parameters used by as much as $95\%$. We provide extensive analysis of results stratified across polyp categories, scanner types, imaging modalities, and focus modes to establish a new benchmark on this challenging, unexplored, large-scale dataset and promote future direction into the use of AI-driven colorectal cancer screening systems.

\begin{figure*}[t]
\begin{center}
  \includegraphics[width=0.95\linewidth]{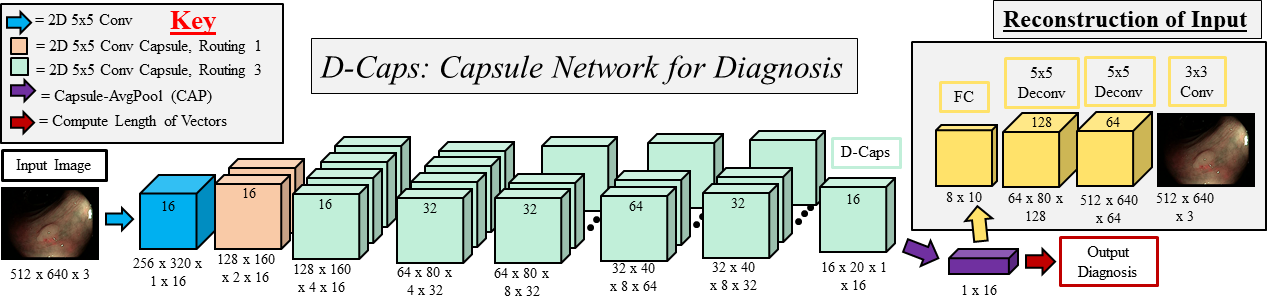}
\end{center}
  \caption{D-Caps: Diagnosis capsule network architecture. Routing 1 or 3 refers to the number of routing iterations performed.}
\label{fig:Network}
\end{figure*}

\section{Related Work}

For computer-aided diagnosis (CAD) studies, colorectal polyp diagnosis is fairly limited. In \cite{KOMINAMI2016643}, a bag-of-features representation was constructed by a hierarchical k-means clustering of scale invariant feature transform (SIFT) descriptors.  These features were then used to train an SVM classifier for classifying hyperplastic polyps vs adenomas. The approach by \cite{Zhang} was the first to incorporate deep learning to diagnose hyperplastic polyps vs adenomas. The authors extracted the first $3$ -- $4$ layers of an Inception-style network trained on ImageNet and Places205 and trained an SVM to classify the extracted deep features. The first end-to-end trained network was used in \cite{Byrnegutjnl}, which employed an AlexNet style network trained from scratch with data augmentation to classify polyps as hyperplastic, adenomas, none, or unsuitable image. Most recently, \cite{CHEN2018568} used a pretrained Inceptionv3 network to classify hyperplastic polyps from adenomas. For comparison in this study, we employ the Inceptionv3 network as our baseline (see section~\ref{sec:data}).

Capsule networks were popularized in \cite{Sabour}, where a small capsule network (CapsNet) comprised of a single layer of convolutional capsules without routing and a single layer of fully-connected capsules with dynamic routing were used to classify digits and small images. The network produced state-of-the-art results on MNIST and relatively strong results on CIFAR-10. Several notable capsule network studies have followed; specific to medical imaging applications, \cite{jimenez2018capsule} employed CapsNet for a number of medical and non-medical tasks, and show capsule networks may generalize better given limited data. To classify brain tumor types, \cite{afshar2018brain} also employed an unmodified CapsNet. In \cite{lalonde2018capsules}, the authors locally constrain the dynamic routing algorithm to allow for deeper networks and share transformation matrices across members of the grid while forming an encoder-decoder style network for pathological lung segmentation. In our proposed work, we follow this style of dynamic routing and transformation matrix sharing, while introducing capsule-average pooling, to allow for our deep network classification architecture D-Caps.

\section{Methods}
The proposed D-Caps is illustrated in Fig ~\ref{fig:Network}. Briefly, input to the network is a $512 \times 640 \times 3$ color image taken during colonoscopy screening. This image is sent through an initial convolutional layer which downsamples the image and extracts the basic low-level feature information (edges and corners). This output is reshaped to be treated as a convolutional capsule with a single capsule type, whose feature vectors are then passed to the first convolutional capsule layer, referred to as the primary capsules and a second capsule type is added. All further layers are convolutional capsule layers with locally connected dynamic routing, until the capsule-average pooling layer and reconstruction sub-network. 

In each capsule layer, there are individual capsules which form a grid. Then, at each layer, there are multiple sets of these grids which form the capsule types. Capsules within a lower layer are referred to as child capsules and in a higher layer being routed to as parent capsules. The locally connected dynamic routing works by forming prediction vectors over a kernel of the child capsules centered at the location of the set of parent capsule types. For every parent capsule at a given $(x,y)$ position, a set of prediction vectors are formed via the multiplication between a locally-defined window and a transformation matrix which is shared across the spatial dimension (but not the capsule type dimension). These transformation matrices act analogous to affine transformation in feature space, allowing for a strong notion of equivariance to input features. Once prediction vectors are formed for a given $(x,y)$ location, and therefore set of parent capsules, the modified dynamic routing algorithm then routes all child capsules to all parents capsules only at that given spatial location.

The \textbf{capsule-average pooling (CAP) layer} computes the spatial average of capsule activation vectors to reduce the dimensionality of the features. Each capsule type computes an element-wise \textit{mean} across the height and width dimensions of the capsule grid, preserving the length of the capsule vectors in each capsule type. Since, in our application, we are computing a binary classification, we have one capsule type in the final convolutional capsule layer, which transforms to a 1D vector of length $k$, in our case $k=16$. More explicitly, if we have $n$ capsule types, each with $h \times w$ grids of capsule vectors of length $a$, we compute
\begin{equation}
p_a^i = \frac{1}{h \times w}\sum_{h}\sum_{w}c_{h,w,a}^i, \forall i \in \{1, 2, ..., n\}.
\label{eq:cap}
\end{equation}
In previous approaches, a fully-connected capsule layer is used to predict the final class-activation vectors. This becomes computationally infeasible with any reasonable sized GPU memory when working with large-scale images and number of classes. By utilizing our CAP layer, we are able to dramatically increase the size of the images we work with beyond the likes of MNIST, CIFAR-10 and smallNORB. The D-Caps architecture shown in Fig.~\ref{fig:Network} contains only $1.3$ million parameters, as compared to $24$ million in Inceptionv3, a relative reduction of $95\%$, while achieving higher performance. 

\textbf{To decide a class score:} the magnitude of each vector is computed, where the longest vector is chosen as the prediction. In the case where multiple images of the same polyp were given, the votes for each images are averaged, weighted by the relative confidence of the vote being cast. Reconstruction of the input is then performed via a dense layer followed by two deconvolutions and a final convolution. The reconstruction serves the purpose of providing a learned inverse mapping from output to input, in order to help preserve a better approximation of the distribution of the input space. Without the inverse mapping, the network will be prone to only learn the most common modes in the training dataset. We show in an ablation study this reconstruction significantly helps the accuracy of our approach, which is not possible with a standard CNN that only represents features as scalars.

\begin{table*}[ht]
\centering
\caption{Classifying \textbf{Hyperplastic vs Adenoma polyps} measured by accuracy (acc), sensitivity (sen), and specificity (spe), where -F and -N denote far and near focus, respectively.}
\label{table:m1}
\begin{tabular}{@{}cc|ccccccccccc@{}}
\toprule
\multicolumn{2}{c}{Method} & All Images & All Polyps & NBI & NBI-F & NBI-N & WL & WL-F & WL-N & Near & Far & \\
\midrule
\multirow{3}{*}{D-Caps} &
    Acc. \% & $\bm{63.66}$ & $\bm{65.53}$ & $\bm{56.69}$ & $53.37$ & $\bm{60.95}$ & $\bm{68.81}$ & $\bm{72.48}$ & $\bm{67.65}$ & $\bm{67.57}$ & $\bm{69.64}$ & \\
     & Sen. \% & $\bm{65.26}$ & $\bm{71.12}$ & $54.23$ & $51.97$ & $\bm{59.74}$ & $\bm{74.06}$ & $\bm{75.63}$ & $\bm{70.86}$ & $\bm{70.19}$ & $\bm{73.62}$ & \\
     & Spe. \% & $\bm{60.00}$ & $\bm{53.79}$ & $\bm{61.98}$ & $\bm{57.14}$ & $\bm{64.29}$ & $\bm{57.38}$ & $\bm{63.79}$ & $\bm{58.49}$ & $\bm{60.66}$ & $\bm{59.02}$ & \\
\midrule
\multirow{3}{*}{Iv3} &
    Acc. \% & $54.28$ & $56.23$ & $52.49$ & $\bm{58.65}$ & $53.33$ & $55.41$ & $55.50$ & $58.33$ & $57.66$ & $58.48$ & \\
     & Sen. \% & $54.83$ & $63.18$ & $\bm{57.69}$ & $\bm{59.21}$ & $56.49$ & $54.89$ & $53.75$ & $58.94$ & $63.35$ & $63.19$ & \\
     & Spe. \% & $53.00$ & $41.67$ & $41.32$ & $\bm{57.14}$ & $44.64$ & $56.56$ & $60.34$ & $56.60$ & $42.62$ & $45.90$ & \\
\bottomrule
\end{tabular}
\end{table*}

\begin{table*}[ht]
\centering
\caption{Classifying \textbf{Hyperplastic vs Adenoma and Serrated polyps} measured by accuracy (acc), sensitivity (sen), and specificity (spe), where -F and -N denote far and near focus, respectively.}
\label{table:m2}
\begin{tabular}{@{}cc|ccccccccccc@{}}
\toprule
\multicolumn{2}{c}{Method} & All Images & All Polyps & NBI & NBI-F & NBI-N & WL & WL-F & WL-N & Near & Far & \\
\midrule
\multirow{3}{*}{D-Caps} &
    Acc. \% & $\bm{59.81}$ & $\bm{60.95}$ & $\bm{60.36}$ & $\bm{60.09}$ & $\bm{63.59}$ & $\bm{54.39}$ & $55.86$ & $\bm{56.67}$ & $\bm{58.52}$ & $\bm{62.01}$ & \\
     & Sen. \% & $\bm{61.39}$ & $\bm{63.19}$ & $\bm{60.00}$ & $\bm{59.24}$ & $\bm{65.22}$ & $\bm{59.21}$ & $64.02$ & $\bm{58.60}$ & $\bm{60.12}$ & $\bm{67.86}$ & \\
     & Spe. \% & $\bm{56.00}$ & $\bm{56.06}$ & $\bm{61.16}$ & $\bm{62.50}$ & $\bm{58.93}$ & $\bm{43.44}$ & $32.76$ & $\bm{50.94}$ & $\bm{54.10}$ & $\bm{45.90}$ & \\
\midrule
\multirow{3}{*}{Iv3} &
    Acc. \% & $51.21$ & $48.10$ & $45.27$ & $51.17$ & $46.54$ & $51.88$ & $\bm{59.01}$ & $50.95$ & $47.60$ & $50.22$ & \\
     & Sen. \% & $53.49$ & $50.35$ & $41.11$ & $50.96$ & $46.58$ & $56.68$ & $\bm{66.46}$ & $\bm{58.60}$ & $51.79$ & $55.36$ & \\
     & Spe. \% & $45.75$ & $43.18$ & $54.55$ & $51.79$ & $46.43$ & $40.98$ & $\bm{37.93}$ & $28.30$ & $36.07$ & $36.07$ & \\
\bottomrule
\end{tabular}
\end{table*}

\begin{table*}[!ht]
\centering
\caption{Classifying \textbf{Hyperplastic vs Serrated polyps} measured by accuracy (acc), sensitivity (sen), and specificity (spe), where -F and -N denote far and near focus, respectively.}
\label{table:m3}
\begin{tabular}{@{}cc|cccccccccc@{}}
\toprule
\multicolumn{2}{c}{Method} & All Images & All Polyps & NBI & NBI-F & NBI-N & WL & WL-F & WL-N & Near & Far \\
\midrule
\multirow{3}{*}{D-Caps} &
    Acc. \% & $\bm{60.91}$ & $\bm{58.04}$ & $\bm{57.85}$ & $\bm{55.00}$ & $\bm{60.00}$ & $\bm{54.14}$ & $\bm{52.63}$ & $\bm{52.54}$ & $\bm{67.21}$ & $\bm{66.67}$ \\
     & Sen. \% & $\bm{65.00}$ & $54.55$ & $70.00$ & $60.00$ & $71.43$ & $\bm{54.55}$ & $\bm{100.00}$ & $\bm{50.00}$ & $57.14$ & $60.00$ \\
     & Spe. \% & $\bm{60.50}$ & $\bm{58.33}$ & $\bm{56.76}$ & $\bm{54.29}$ & $\bm{57.58}$ & $\bm{54.10}$ & $\bm{49.06}$ & $\bm{52.83}$ & $\bm{68.52}$ & $\bm{67.21}$ \\
\midrule
\multirow{3}{*}{Iv3} &
    Acc. \% & $51.45$ & $40.54$ & $45.63$ & $44.90$ & $50.00$ & $48.08$ & $41.86$ & $45.00$ & $40.00$ & $39.62$ \\
     & Sen. \% & $63.64$ & $\bm{66.67}$ & $\bm{83.33}$ & $\bm{66.67}$ & $\bm{100.00}$ & $16.67$ & $66.67$ & $33.33$ & $\bm{100.00}$ & $\bm{66.67}$ \\
     & Spe. \% & $50.62$ & $39.05$ & $43.30$ & $43.48$ & $46.81$ & $50.00$ & $40.00$ & $45.95$ & $36.17$ & $38.00$ \\
\bottomrule
\end{tabular}
\end{table*}

\section{Datasets \& Experiments}\label{sec:data}

Experiments were performed on a Mayo Polyp dataset, collected at the Mayo Clinic, Jacksonville by \cite{WALLACE20141072} with an institutional review board approval. A total of 552 patients were included in this study with $963$ polyps collected. Polyps were collected from both standard and dual-focus colonoscopes. The dual-focus colonoscope contains near and far modes for both white light (WL) and narrow-band imaging (NBI) settings, referred to as WL-N, WL-F, NBI-N, and NBI-F, respectively. Challenging images of each polyp type are chosen by expert interpreters (one per imaging type). 

Three sets of experiments were conducted using stratified 10-fold cross validation. In the first set, images were split into two categories, hyperplastics and adenomas (with serrated adenomas excluded). In the second set, the serrated adenomas were included in the adenoma class. In the third set, images were split between hyperplastics and serrated adenomas with the adenoma images excluded. The results of these experiments are presented in the following section. Additionally, we conducted three rounds of ablation experiments with results presented at the polyp level: i) varying the amount of dynamic routing iterations performed inside D-Caps, ii) removing the reconstruction regularization sub-network, and iii) evaluating D-Caps performance on an `ideal' subset of 95 NBI-N images selected by participating physicians for homogeneity to see performance in more ideal cases of using near-focus NBI colonoscopes with good scale/centering on polyps.

All networks were trained and tested on a single Titan X GPU using the Keras and TensorFlow frameworks. Both Inceptionv3 and D-Caps were trained from scratch using the Adam optimizer at its default settings. A batch size of $8$ was used for Inceptionv3 and $4$ for D-Caps due to memory constraints on capsules. The loss function for all networks was a binary cross-entropy. All code for reproducing experiments are made publicly available.

\section{Results \& Discussion}
The results of the three sets of experiments in presented in Tables \ref{table:m1} - \ref{table:m3}. For all experiments, we present results at several levels of analysis: All Images presents results for every image present in the dataset, while all other results are a weighted average taken across all votes for a given polyp (and imaging modality) to give a final diagnosis score. Looking at the All Polyps columns, we can see D-Caps outperforms Inceptionv3 in terms of relative accuracy increases of $17\%$, $27\%$, and $43\%$ for experiments $1$ -- $3$ (of increasing difficulty) respectively. 

In our routing iteration ablation experiment, we obtained  $50.61\%$, $65.53\%$, $45.97\%$, and $50.86\%$ accuracy at the polyp level for $2$, $3$, $4$, and $5$ routing iterations respectively. Removing the reconstruction sub-network obtained $56\%$, $50\%$, and $55\%$ accuracy at the for experiments $1$ -- $3$ respectively, an average $8\%$ decrease. Lastly on the ideal subset of physician chosen images, we obtained an accuracy of $82\%$ for hyperplastic vs adenoma. These experiments show the dynamic routing and reconstruction both contribute to the overall performance of our model, while the latter experiment provides strong evidence that with further improvements in both capsule network algorithms and screening technology, AI-driven approaches can prove viable for raising optical biopsy techniques to clinical practice standards. Our work provides an important baseline for future studies on the extremely challenging Mayo Polyp dataset, and contributes further evidence that given limited data with high intra-class variation, capsule networks can significantly outperform deep CNNs.

\bibliographystyle{IEEEbib}
\bibliography{main}

\end{document}